\documentclass[doublecol,figures,reply]{epl2}

\title{Reply to the Comment by S.E. Sebastian and N. Harrison}
\shorttitle{Field dependence of the quantum ground state in SrCu$_2$(BO$_3$)$_2$} %Insert here a short version of the title if it exceeds 70 characters

\author{
F. L\'{e}vy\inst{1} \and I. Sheikin\inst{1} \and C. Berthier\inst{1} \and M.
Horvati\'c\inst{1} \and M. Takigawa\inst{2} } \shortauthor{F. L\'{e}vy \etal}

\institute{
 \inst{1} Grenoble High Magnetic Field Laboratory
 (GHMFL) - CNRS, BP 166, 38042 Grenoble Cedex 09, France \\
 \inst{2}
 Institute for Solid State Physics, University of Tokyo,
 Kashiwanoha, Kashiwa, 277-8581, Japan \\
}
 {Quantized spin models} 
{Magnetic phase boundaries}

\begin{document}

\maketitle

In their comment, Sebastian and Harrison (SH) suggest that our
torque and force measurements in SrCu$_2$(BO$_3$)$_2$
\cite{Levy_08} are incorrect due to a non-linear regime in the
cantilever operation and deny the importance of
Dzyaloshinski-Moriya (DM) interaction in torque measurements. In
this reply, we show that their arguments are incorrect and why
neither the torque measurements \cite{Levy_08,Sebastian_08} nor
the magnetization measurements in pulsed magnetic field
\cite{Onizuka_00, Jorge_05} can give access to the field
dependence of the longitudinal magnetization at thermodynamic
equilibrium.

First of all, any non-linearity is perfectly excluded in all our
measurements. Our largest signal reached only ~10\,\% of the whole
linear response range, and the variation of the capacitance has
always been smaller than 1\,\% of the zero field capacitance. This
means that the deflection of the cantilever was smaller that
10$^{-4}$ rad, which is comparable to the numbers given in the
comment (0.5$\times$10$^{-4}$ rad). Next, the angle $\theta$
between the $c$-axis of the sample and the applied magnetic field
$H$ in our experiment was smaller than 0.4$^{\circ}$, which is
much smaller than the 2$^{\circ}$ mentioned in \cite{Sebastian_08}
and comparable to the $\ll 1^{\circ}$ declared for the experiments
reported in the comment.

SH base their interpretation of the torque results on the
assumption that DM effects are negligible, as their data follow
closely those obtained by inductive method in pulsed magnetic
field, which couples only to the total magnetization $M_{z}$. This
statement is obviously wrong since in the gapped phase below 16~T
the pulsed field measurements correctly record zero magnetization,
while the torque measurements \emph{do} record a non-zero signal -
that is the signal due to DM interaction. Indeed, Ref.
\cite{Sebastian_08} explicitly mentions that at low fields the
data have been corrected (put to zero) by hand, and in the Fig.~1
of the comment it is obvious that below 16~T there is an important
variation of the torque signal. Furthermore, in the same figure
the steep jump towards the 1/8 plateau is obviously \emph{not} the
same in the torque and in the pulse field data. So the main
assumption of SH is contradicted by their own data. Finally,
Fig.~3 in Ref. \cite{Onizuka_00} shows that below 30~T the
magnetization measurements do \emph{not} scale with the $g$-tensor
anisotropy, at variance with their claims.

SH criticize the large jump in the magnetization curve observed in
our force measurements \cite{Levy_08}, in which the deviation of
the cantilever is dominated by the force  $F = \frac{dH}{dz}M_z$.
The variation of $M_z$ during this "jump" indeed corresponds to
the coexistence of the uniform magnetization phase and the 1/8
plateau phase as already shown by copper NMR \cite{Kodama_02}.
\begin{figure}[t]
\begin{center}
\includegraphics[width=\linewidth]{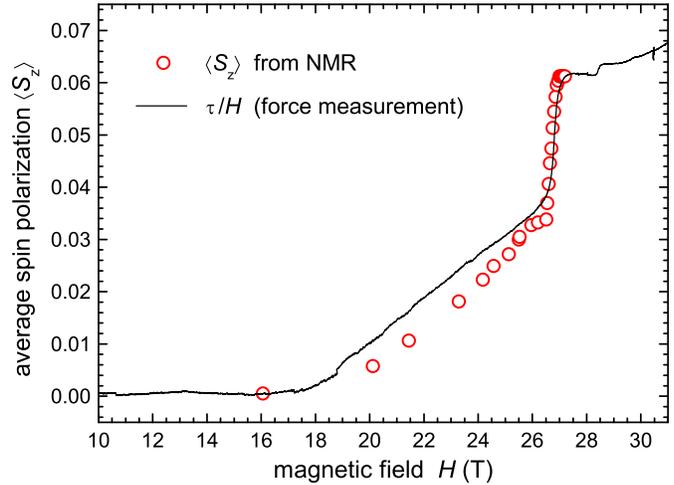}
\caption{\textbf{Magnetization from NMR and force measurements}
The field dependance of magnetization, and in particular the jump
towards the 1/8 plateau, recorded by our force measurement
\cite{Levy_08} is in excellent agreement with values determined
from the NMR data of Ref. \cite{Kodama_02} as explained in the
text.} \label{fig1}
\end{center}
\end{figure}
Kodama \emph{et al.} have shown that up to 26.5\,T the system is
in a uniform phase in which all copper electronic spins bear
identical magnetization $g\mu_{\rm{B}} S_{z}$. In this phase,
 $S_{z}$ values have been accurately determined by NMR and found to increase up
to the value of 0.034 at 26.5\,T (Fig.~4A of Ref.
\cite{Kodama_02}). Between 26.5\,T and 27\,T, the increase of the
volume fraction $x(H)$ of the plateau phase as a function of $H$
has also been determined by NMR (Fig.~4C of Ref.
\cite{Kodama_02}). One can thus easily deduce the field dependence
of the average bulk magnetization $\langle S_{z} \rangle =
0.034\,(1-x) + 0.0625\,x$, which is reported in Figure~\ref{fig1}.
Clearly, the agreement between the force measurement and the NMR
data is excellent, meaning that the rapid variation of $M_z(H)$ in
the field range 26.5\,--\,27~T \emph{does} correspond to the
physics of the system. This clearly indicates that all previously
reported measurements, whatever using torque technique in static
field or flux integration in pulsed field, were unable to give the
correct equilibrium $M_z(H)$ dependence.

In the following, we explain the possible origin of discrepancies
between different techniques. In pulsed magnetic field
measurements, during the fast increase of the field the energy
levels of the lower triplet states are lowered towards the singlet
level, but their populations have not enough time to relax to its
thermal equilibrium value (adiabatic process). The effective
temperature describing this non-equilibrium level populations is
thus much lower than that of the experiment. This explains the
observation of 1/8 magnetization plateau in pulsed magnetic field
experiments performed at 1.4\,K \cite{Onizuka_00}, 1.6\,K and
0.6\,K \cite{Jorge_05}, although this phase is \emph{not} stable
above 0.55\,K. This also implies that the values of $M_z$ in these
experiments are incorrect.

Let us come to the torque measurements, and why the results of
different groups can differ. There are three contributions to the
torque applied on the cantilever. The first one is well known and
 related to the anisotropy of the
diagonal part of the $g$-tensor. The second one is due to the
field gradient ($dH/dz \neq 0$) if the sample is not rigorously at
the center of the field. The third one is related to the presence
of the DM interaction and the staggered $g$-tensor in the system.
They induce a staggered transverse magnetization, but also a
\emph{uniform} transverse magnetization contributing to the torque
\cite{Miyahara_07}. This contribution vanishes at zero $\theta$,
but is otherwise always present in experimental results
\cite{Levy_08}. The interplay between the three contributions to
the torque is very dependent on the position $z$ of the sample
with respect to the center of the field and the angle $\theta$,
which can be varied independently. We first focus on the first two
terms. For small values of $\theta$, the torque divided by the
magnetic field can be written as:  $\tau/H = [(g_{c}-g_{a})/g_{c}
\sin 2\theta +~l/H~dH/dz ] M_{z}$,  where $l$ is the length of the
cantilever, $g_c$~= 2.28 and $g_a$~= 2.05 \cite{Nojiri_99}. Close
to the center the field profile is parabolic, so that $dH/dz =
z\,d^2H/dz^2$. Clearly one can find a position $z_0$ where the two
terms cancel: $ z_0 = -~2\theta/l
~(g_{c}-g_{a})/g_{c}[d^2H/Hdz^2]^{-1}.$  For typical values of
$\frac{d^2H}{Hdz^2}$~=~$-$50\,ppm/mm$^2$ and $l$~=~5\,mm, this
leads to $z_0$\,(mm)~=~14~$\theta(^{\circ}$), e.g. $z_0$~= +~6\,mm
for $\theta$~= 0.4$^{\circ}$. That is, depending on the position
of the sample, one can not simply predict the relative magnitude
and sign of the contribution proportional to $M_z$ and that due to
the transverse magnetization induced by the DM interaction. This
can well explain why our torque measurements exhibit a negative
slope within the plateau, while those of SH exhibit a positive
slope. Unfortunately, there is no calculation yet of the
transverse uniform magnetization generated by the DM and the
staggered $g$ tensor on a Shastry-Sutherland lattice, in presence
of a superstructure of the magnetization corresponding to a
plateau at $M_z/M_{saturation}$~= 1/8~\cite{Kodama_02} or
1/9~\cite{Dorier_08}.

One puzzling feature in our results, however, is that the signal
amplitudes obtained in the force and in the torque measurements
are comparable, although in principle one would expect the
contribution of the DM interaction to be much smaller
\cite{Miyahara_07}.  Since we have not done a systematic study as
a function of the angle $\theta$ and the position $z$,  we cannot
disentangle the relative contributions of the longitudinal
magnetization and the transverse one. Furthermore, it has been
shown recently \cite{Manama_09} that in  frustrated spin ladders
the DM interaction can give a huge contribution to the torque when
the angle between the magnetic field and the DM vector is small.
In SrCu$_2$(BO$_3$)$_2$ the interdimer DM vector is nearly
parallel to the c-axis \cite{Cheng_07}, and may thus be a possible
source of a strong torque signal. As regards our force
measurement, considering their agreement with the NMR data, they
appear to give a correct field dependence of the magnetization, at
least up to the 1/8 plateau. The theoretical prediction for the
absolute value of the magnetization in the plateaus in
SrCu$_2$(BO$_3$)$_2$ is currently controversial
\cite{Sebastian_08,Dorier_08, Abendschein_08}, which makes their
experimental determination highly desirable. This requires a
special experimental setup to minimize as much as possible any
torque contribution with respect to the effect of force due to the
field gradient applied on the sample \cite{Sakikabara_94}.

%\acknowledgments  This work was supported by the the French ANR
%grant 06-BLAN-0111 and Grant-in-Aids for Scientific Research (Nos.
%16076204 and 19052004) from the MEXT Japan.

\end{document}